\documentclass[12pt]{article}

\usepackage{graphicx}
\usepackage{float}
\usepackage{cite}
\usepackage{amsfonts}
\usepackage{amssymb}
\usepackage{amsmath}

\usepackage{xcolor}
\usepackage{cancel}
\usepackage[normalem]{ulem}

\def\gtwid{\mathrel{\raise.3ex\hbox{$>$\kern-.75em\lower1ex\hbox{$\sim$}}}}
\def\ltwid{\mathrel{\raise.3ex\hbox{$<$\kern-.75em\lower1ex\hbox{$\sim$}}}}
\def\square{\kern1pt\vbox{\hrule height 1.2pt\hbox{\vrule width 1.2pt\hskip 3pt
   \vbox{\vskip 6pt}\hskip 3pt\vrule width 0.6pt}\hrule height 0.6pt}\kern1pt}

\begin{document}

\begin{titlepage}

\begin{flushright}
UFIFT-QG-23-08
\end{flushright}

\vskip 0.2cm

\begin{center}
{\bf Explaining Large Electromagnetic Logarithms \\
from Loops of Inflationary Gravitons}
\end{center}

\vskip 0.2cm

\begin{center}
D. Glavan$^{1*}$, S. P. Miao$^{2\star}$, T. Prokopec$^{3\dagger}$ and
R. P. Woodard$^{4\ddagger}$
\end{center}
\vspace{0.cm}
\begin{center}
\it{$^{1}$ CEICO, Institute of Physics of the Czech Academy of Sciences (FZU), \\
Na Slovance 1999/2, 182 21 Prague 8, CZECH REPUBLIC}
\end{center}
\vspace{-0.3cm}
\begin{center}
\it{$^{2}$ Department of Physics, National Cheng Kung University, \\
No. 1 University Road, Tainan City 70101, TAIWAN}
\end{center}
\vspace{-0.3cm}
\begin{center}
\it{$^{3}$ Institute for Theoretical Physics, Spinoza Institute \& EMME$\Phi$, \\
Utrecht University, Postbus 80.195, 3508 TD Utrecht, THE NETHERLANDS}
\end{center}
\vspace{-0.3cm}
\begin{center}
\it{$^{4}$ Department of Physics, University of Florida,\\
Gainesville, FL 32611, UNITED STATES}
\end{center}

\vspace{0.2cm}

\begin{center}
ABSTRACT
\end{center}
Recent progress on nonlinear sigma models on de Sitter background has
permitted the resummation of large inflationary logarithms by combining
a variant of Starobinsky's stochastic formalism with a variant of the
renormalization group. We reconsider single graviton loop corrections 
to the photon wave function, and to the Coulomb potential, in light of
these developments. Neither of the two 1-loop results have a stochastic 
explanation, however, the flow of a curvature-dependent field strength 
renormalization explains their factors of $\ln(a)$. We speculate that 
the factor of $\ln(Hr)$ in the Coulomb potential should not be 
considered as a leading logarithm effect. 

\begin{flushleft}
\end{flushleft}

\vspace{0.2cm}

\begin{flushleft}
$^{*}$ e-mail:glavan@fzu.cz \\
$^{\star}$ e-mail: spmiao5@mail.ncku.edu.tw \\
$^{\dagger}$ e-mail: T.Prokopec@uu.nl \\
$^{\ddagger}$ e-mail: woodard@phys.ufl.edu
\end{flushleft}

\end{titlepage}

\section{Introduction}

Cosmology is characterized by scale factor $a(t)$, Hubble parameter $H(t)$
and first slow roll parameter $\epsilon(t)$,
\begin{equation}
ds^2 = -dt^2 + a^2(t) d\vec{x} \!\cdot\! d\vec{x} \qquad \Longrightarrow \qquad
H(t) \equiv \frac{\dot{a}}{a} \quad , \quad \epsilon(t) \equiv -\frac{\dot{H}}{H^2}
\; . \label{geometry}
\end{equation}
Inflation is the special case for which both the first and second time derivatives
of the scale factor are positive ($H(t) > 0$ and $0 \leq \epsilon(t) < 1$). It is
the accelerated expansion of inflation which produces the primordial spectra of
scalars \cite{Mukhanov:1981xt} and gravitons \cite{Starobinsky:1979ty} by 
ripping these quanta out of the vacuum. 

At some level these quanta must interact with themselves and with other particles.
These interactions can change single particle kinematics and long range forces,
and one might expect that the changes grow because more and more quanta are ripped
out of the vacuum as time progresses. For example, a single loop of gravitons on 
de Sitter background ($\epsilon(t) = 0$) corrects the electric field strength of
a plane wave photon \cite{Wang:2014tza} and the Coulomb potential of a point
charge \cite{Glavan:2013jca} to,
\begin{eqnarray}
F^{0i}(t,\vec{x}) &\!\!\! = \!\!\!& F^{0i}_{\rm tree}(t,\vec{x}) \Biggl\{ 1 + 
\frac{2 G H^2}{\pi} \ln(a) + O(G^2) \Biggr\} \; , \qquad \label{photon} \\
\Phi(t,r) &\!\!\! = \!\!\!& \frac{Q}{4\pi a r} \Biggl\{1 + \frac{2 G}{3\pi a^2 r^2} 
+ \frac{2 G H^2}{\pi} \ln(a H r) + O(G^2) \Biggr\} \; . \qquad \label{Coulomb}
\end{eqnarray}
Similar results have been reported for fermions \cite{Miao:2006gj}, for massless,
minimally coupled scalars \cite{Glavan:2021adm}, and for gravitons 
\cite{Tan:2021lza,Tan:2022xpn}. 

A fascinating aspect of these results is that they continue to grow for as long 
as inflaton persists. For sufficient inflation, the factors of $\ln[a(t)]$ must 
eventually overwhelm the loop-counting parameter $G H^2$ causing perturbation
theory to break down. Evolving past this point requires a nonperturbative 
resummation technique of the sort recently developed for nonlinear sigma models
on de Sitter background \cite{Miao:2021gic,Woodard:2023rqo,Litos:2023nvj}. The 
technique combines a variant of Starobinsky's stochastic formalism 
\cite{Starobinsky:1986fx,Starobinsky:1994bd}, based on curvature-dependent 
effective potentials, with a variant of the renormalization group, based on the 
subset of counterterms which can be viewed as curvature-dependent renormalizations 
of parameters in the bare theory. The latter part of the technique is not 
encountered in renormalizable matter theories, where the curvature-independent 
renormalization group explains large secular logarithms~\cite{Glavan:2023lvw}. 
Even better, the technique can be generalized to a arbitrary cosmological background 
(\ref{geometry}) which has undergone primordial inflation \cite{Kasdagli:2023nzj}, 
and applying it transmits inflationary effects to late times \cite{Woodard:2023cqi}.

It seems entirely possible to generalize this technique from nonlinear sigma 
models to quantum gravity. The first step has been taken by using a variant of 
the renormalization group to explain the large logarithm in the 1-graviton loop
correction to the exchange potential of a massless, minimally coupled scalar 
\cite{Glavan:2021adm}. The purpose of this paper is to do the same for the 
1-graviton loop corrections (\ref{photon}-\ref{Coulomb}) to electrodynamics. In 
section 2 we review the exact calculation. Section 3 uses the renormalization 
group to explain the factors of $\ln[a(t)]$ in both results. We do not believe 
there is any curvature-dependent effective potential for this system, and we 
suspect that the factor of $\ln(H r)$ in (\ref{Coulomb}) is not a leading 
logarithm effect. The case for that is made in section 4. Our conclusions 
comprise section 5.

\section{The Exact Calculation}

The purpose of this section is to review the exact calculation of the 
1-graviton loop contribution to the vacuum polarization $i[\mbox{}^{\mu}
\Pi^{\nu}](x;x')$ \cite{Leonard:2013xsa} from which the results 
(\ref{photon}-\ref{Coulomb}) were derived \cite{Wang:2014tza,Glavan:2013jca}. 
These results were obtained by perturbatively solving the quantum-corrected 
Maxwell equation,
\begin{equation}
\partial_{\nu} \Bigl[ \sqrt{-g} \, g^{\nu\rho} g^{\sigma\mu} F_{\rho\sigma}(x)
\Bigr] + \int \!\! d^4x' \Bigl[\mbox{}^{\mu} \Pi^{\nu}\Bigr](x;x') A_{\nu}(x')
= J^{\mu}(x) \; , \label{QMax}
\end{equation}
where $F_{\mu\nu} \equiv \partial_{\mu} A_{\nu} - \partial_{\nu} A_{\mu}$ is
the field strength tensor and $J^{\mu}$ is the current density. We begin by
explaining how the vacuum polarization is represented and why de Sitter 
breaking is unavoidable. Next the counterterms are given. The section closes
by giving the structure functions and isolating those terms which are 
responsible for the large logarithms in (\ref{photon}-\ref{Coulomb}). Throughout
we employ conformal coordinates (based on $d\eta \equiv dt/a(t)$) so that the
de Sitter metric $g_{\mu\nu} = a^2 \eta_{\mu\nu}$ is proportional to Minkowski
metric of flat space.

Of the ten graviton loops which have so far been evaluated on de Sitter background
\cite{Tsamis:1996qm,Tsamis:1996qk,Tsamis:2005je,Miao:2005am,Kahya:2007bc,
Miao:2012bj,Leonard:2013xsa,Glavan:2015ura,Miao:2017vly,Glavan:2020gal}, 
\footnote{See also the computation of graviton corrections to massless, conformally
coupled scalars \cite{Boran:2014xpa,Boran:2017fsx} which disagrees with our result
\cite{Glavan:2020gal}.} all but one of them \cite{Glavan:2015ura} used the simplest 
gauge \cite{Tsamis:1992xa,Woodard:2004ut}. The great thing about the dimensionally
regulated (spacetime dimension $D$) propagator in this gauge is that it consists of 
three scalar propagators (with masses $M_A^2 = 0$, $M_B^2 = (D-2) H^2$ and $M_C^2 = 
2 (D-3) H^2$) multiplied by constant tensor factors which are formed using the 
Minkowski metric $\eta_{\mu\nu}$ and $\delta^0_{~\mu}$ (with $\overline{\eta}_{\mu\nu} 
\equiv \eta_{\mu\nu} + \delta^0_{~\mu} \delta^0_{~\nu}$),
\begin{eqnarray}
i \Bigl[\mbox{}_{\mu\nu} \Delta_{\rho\sigma}\Bigr](x;x') = \sum_{I = A,B,C} 
i\Delta_I(x;x') \!\times\! \Bigl[\mbox{}_{\mu\nu} T^I_{\rho\sigma}\Bigr] \; , 
\qquad \label{simpprop} \\
\Bigl[\mbox{}_{\mu\nu} T^A_{\rho\sigma}\Bigr] = 2 \overline{\eta}_{\mu (\rho}
\overline{\eta}_{\sigma) \nu} - \frac{2 \overline{\eta}_{\mu\nu}
\overline{\eta}_{\rho\sigma}}{D \!-\! 3}  \;\; , \;\; \Bigl[\mbox{}_{\mu\nu} 
T^B_{\rho\sigma}\Bigr] = -4 \delta^0_{~(\mu} \overline{\eta}_{\nu) (\rho}
\delta^0_{~\sigma)} \; , \qquad \label{TAB} \\
\Bigl[\mbox{}_{\mu\nu} T^C_{\rho\sigma}\Bigr] = \frac{2 [ \overline{\eta}_{\mu\nu} 
+ (D \!-\! 3) \delta^0_{~\mu} \delta^0_{~\nu}] [\overline{\eta}_{\rho\sigma} + 
(D \!-\! 3) \delta^0_{~\rho} \delta^0_{~\sigma}]}{(D \!-\! 3) (D \!-\! 2)} \; . 
\qquad \label{TC}
\end{eqnarray}
Another huge advantage of this gauge is that the $D=4$ dimensional limits
of the three scalar propagators are simple,
\begin{eqnarray}
i\Delta_A(x;x') & \longrightarrow & \frac1{4 \pi^2} \Bigl[ \frac1{a a' \Delta x^2}
- \frac{H^2}{2} \ln\Bigl( \frac14 H^2 \Delta x^2\Bigr)\Bigr] \; , \label{D4propA} \\
i\Delta_B(x;x') & \longrightarrow & i \Delta_C(x;x') \longrightarrow \frac1{4 \pi^2} 
\frac1{a a' \Delta x^2} \; , \qquad \label{D4propBC}
\end{eqnarray}
where $\Delta x^2= -(|\eta-\eta'|-i\epsilon)^2+\|\vec x-\vec x'\|^2$,
with $\epsilon>0$ infinitesimal.

Although the propagator (\ref{simpprop}-\ref{TC}) is the easiest to use, this gauge
does break de Sitter invariance, which means that noninvariant counterterms can and 
do occur. The inevitability of de Sitter breaking for the graviton propagator on de
Sitter background has been a contentious issue for decades \cite{Allen:1986tt,
Hawking:2000ee,Higuchi:2001uv,Higuchi:2002sc,Miao:2011fc,Higuchi:2011vw,Miao:2011ng,
Morrison:2013rqa,Miao:2013isa}. However, the presence of noninvariant counterterms
seems to have been settled by the computation of the vacuum polarization in a general 
class of de Sitter invariant gauges \cite{Mora:2012zi}. In spite of the de Sitter
invariant gauge, noninvariant counterterms still arise due to the unavoidable 
breaking in the time-ordered interactions \cite{Glavan:2015ura}. So we will just
go ahead with the result \cite{Leonard:2013xsa} derived in the simplest gauge.

General relativity plus Maxwell is not perturbatively renormalizable 
\cite{Deser:1974zzd,Deser:1974cz}, however, the 1PI (one-particle-irreducible)
$n$-point functions of any quantum field theory can be renormalized, order-by-order
in perturbation theory, using BPHZ (Bogoliubov, Parasiuk \cite{Bogoliubov:1957gp}, 
Hepp \cite{Hepp:1966eg} and Zimmermann \cite{Zimmermann:1968mu,Zimmermann:1969jj})
counterterms. The ones needed to renormalize the 1-loop vacuum polarization on
de Sitter background are \cite{Leonard:2013xsa,Glavan:2015ura},
\begin{eqnarray}
\lefteqn{\Delta \mathcal{L} = \Delta C H^2 F_{ij} F_{k\ell} g^{ik} g^{j\ell} 
\sqrt{-g} + \overline{C} H^2 F_{\mu\nu} F_{\rho\sigma} g^{\mu\rho} g^{\nu\sigma} 
\sqrt{-g} } \nonumber \\
& & \hspace{6cm} + C_4 D_{\alpha} F_{\mu\nu} D_{\beta} F_{\rho\sigma} 
g^{\alpha\beta} g^{\mu\rho} g^{\nu\sigma} \sqrt{-g} \; , \qquad \label{cterms}
\end{eqnarray}
where $D_{\alpha}$ represents the covariant derivative operator. In the simplest
gauge the divergent coefficients are \cite{Leonard:2013xsa},
\begin{equation}
\Delta C = -1 \!\times\! \frac{\kappa^2 \mu^{D-4}}{16 \pi^2 (D \!-\! 4)} \;\; , \;\;
\overline{C} = \frac76 \!\times\! \frac{\kappa^2 \mu^{D-4}}{16 \pi^2 (D \!-\! 4)} 
\;\; , \;\; C_4 = \frac16 \!\times\! \frac{\kappa^2 \mu^{D-4}}{16 \pi^2 (D \!-\! 4)} 
\; , \label{Cterms}
\end{equation}
where $\kappa^2 \equiv 16 \pi G$ is the loop-counting parameter of quantum 
gravity and $\mu$ is the mass scale of dimensional regularization.

Owing to the unavoidable breaking of de Sitter invariance, the vacuum polarization
requires two structure functions \cite{Leonard:2012si}. Various representations
are possible \cite{Leonard:2012ex}, of which we chose the one first employed 
for the vacuum polarization induced by scalar quantum electrodynamics
\cite{Prokopec:2002jn,Prokopec:2002uw},
\begin{eqnarray}
\lefteqn{i \Bigl[\mbox{}^{\mu} \Pi^{\nu}\Bigr](x;x') = \Bigl[ \eta^{\mu\nu} 
\eta^{\rho\sigma} - \eta^{\mu\sigma} \eta^{\nu\rho} \Bigr] \partial_{\rho} 
\partial'_{\sigma} F(x;x') } \nonumber \\
& & \hspace{6.2cm} + \Bigl[ \overline{\eta}^{\mu\nu} \overline{\eta}^{\rho\sigma} 
- \overline{\eta}^{\mu\sigma} \overline{\eta}^{\nu\rho} \Bigr] \partial_{\rho} 
\partial'_{\sigma} G(x;x') \; . \qquad \label{VPrep}
\end{eqnarray}
We employed the Schwinger-Keldysh formalism \cite{Schwinger:1960qe,
Mahanthappa:1962ex,Bakshi:1962dv,Bakshi:1963bn,Keldysh:1964ud,Chou:1984es,
Jordan:1986ug,Calzetta:1986ey,Ford:2004wc} in order to keep the effective field 
equations real and causal. With a convenient choice of the finite parts of
$\Delta C$, $\overline{C}$ and $C_4$, the Schwinger-Keldysh structure functions
are \cite{Glavan:2013jca},
\begin{eqnarray}
\lefteqn{i F(x;x') = \frac{\kappa^2 H^2}{8 \pi^2} \Biggl\{-\ln\Bigl(\frac{\mu a}{2 H}\Bigr)
- \frac{\partial_0}{3 a H} + \frac{\ln(\frac{\mu a}{2 H})}{3 H^2} \partial_{\mu}
\frac1{a^2} \partial^{\mu} \Biggr\} \delta^4(x \!-\! x') } \nonumber \\
& & \hspace{-0.5cm} + \frac{\kappa^2 \partial^6}{384 \pi^3 a a'} \Biggl\{ 
\theta(\Delta \eta \!-\! \Delta r) \Bigl[ \ln[H^2 (\Delta \eta^2 \!-\! \Delta r^2)] 
- 1 \Bigr] \Biggr\} - \frac{\kappa^2 H^2}{128 \pi^3} \Biggl\{ \Bigl[\partial^4 \!+\!
4 \partial^2 \partial_0^2\Bigr] \nonumber \\
& & \hspace{0.7cm} \times \Bigl[\theta(\Delta \eta \!-\! \Delta r) \ln[H^2 (\Delta 
\eta^2 \!-\! \Delta r^2)] \Bigr] - \Bigl[\partial^4 \!-\! 4 \partial^2 \partial_0^2
\Bigr] \theta(\Delta \eta \!-\! \Delta r) \Biggr\} \; , \qquad \label{Fdef} \\
\lefteqn{i G(x;x') = \frac{\kappa^2 H^2}{6 \pi^2} 
\ln\Bigl( \frac{\mu a}{2 H}\Bigr)
\delta^4(x \!-\! x') } \nonumber \\
& & \hspace{2.7cm} + \frac{\kappa^2 H^2 \partial^4}{96 \pi^3} \Biggl\{\theta(\Delta \eta 
\!-\! \Delta r) \Bigl[ \ln \Big(H^2 (\Delta \eta^2 \!-\! \Delta r^2) \Bigr) - 1 \Bigr] \Biggr\} 
\; , \qquad \label{Gdef}
\end{eqnarray}
where $\Delta \eta \equiv \eta - \eta'$ and $\Delta r \equiv \Vert \vec{x} -
\vec{x}'\Vert$. The flat space result \cite{Leonard:2012fs} is recovered by the
terms which contain no net factors of $H$. The terms which contain factors of $H$
represent the new, de Sitter corrections which represent inflationary particle
production.

It remains to comment on the gauge issue. Any quantity with a graviton propagtor,
such as the 1-graviton loop contribution to the vacuum polarization, is liable to
depend on the gauge fixing function. This dependence is easy to quantify in the
flat space result \cite{Leonard:2012fs} and it must therefore be present at least
in the flat space limit of the de Sitter results we have just presented. 
Presumably there is also gauge dependence in the new, de Sitter contributions
\cite{Glavan:2015ura}. Eliminating this gauge dependence is an important problem
for the physical interpretation of results such as (\ref{photon}-\ref{Coulomb}),
and a procedure has been developed for accomplishing this which works in flat
space \cite{Miao:2017feh,Katuwal:2020rkv} and is being generalized to de Sitter
\cite{Glavan:2021adm,Glavan:2023}. However, the issue of gauge dependence has no 
relevance for the study we are making here, of how to explain the large logarithms 
which occur in a specific gauge.

\section{Renormalization Group Explanation}

The purpose of this section is to show how the factors of $\ln(a)$ in expressions
(\ref{photon}-\ref{Coulomb}) can be explained as the renormalization group flow
of a curvature-dependent renormalization of the electromagnetic field strength. 
We accordingly identify the appropriate counterterm and compute the associated
gamma function. Then the Callan-Symanzik equation for Green's functions is written 
down.

The structure functions (\ref{Fdef}) and (\ref{Gdef}) include all information
about 1-graviton loop corrections to the linearized Maxwell equation (\ref{QMax}).
However, the factors of $\ln(a)$ and $\ln(Hr)$ evident in expressions
(\ref{photon}-\ref{Coulomb}) derive from just two terms in $F(x;x')$,\footnote{
For the Coulomb potential, see equation (30) of \cite{Glavan:2013jca}; for the 
photon field strength, see Table~1 of \cite{Wang:2014tza}.}
\begin{eqnarray}
F(x;x') & \longrightarrow & -\frac{\kappa^2 H^2}{8 \pi^2}
\ln\Bigl( \frac{\mu a}{2 H}\Bigr) \delta^4(x \!-\! x') \qquad \nonumber \\
& & \hspace{1.8cm} - \frac{\kappa^2 H^2 \partial^4}{128 \pi^3} \Biggl\{
\theta(\Delta \eta \!-\! \Delta r) \ln\Bigl[H^2 (\Delta \eta^2 \!-\! 
\Delta r^2)\Bigr] \Biggr\} \; , \qquad \label{Fimp} \\
G(x;x') & \longrightarrow & 0 \; . \label{Gimp}
\end{eqnarray}
The factors of $\ln(a)$ in (\ref{photon}-\ref{Coulomb}) come entirely from 
the local term on the first line of (\ref{Fimp}), whereas it is the nonlocal
term on the last line of (\ref{Fimp}) which produces the factor of $\ln(Hr)$
in the Coulomb potential.

The Renormalization Group is associated with the dependence on the dimensional
regularization mass scale $\mu$ which enters through the coefficients 
(\ref{Cterms}) of the counterterms (\ref{cterms}). To understand how this scale
affects the structure functions $F(x;x')$ and $G(x;x')$ we exploit conformal 
coordinates to exhibit the scale factors, and we expand the covariant derivatives 
of the $C_4$ counterterm so that they give ordinary derivatives plus terms which 
can be combined with the $\Delta C$ and $\overline{C}$ counterterms,
\begin{eqnarray}
\lefteqn{\Delta \mathcal{L} = \Bigl[\Delta C - (D \!-\! 6) C_4\Bigr] a^{D-4} H^2
F_{ij} F_{ij} } \nonumber \\
& & \hspace{1.6cm} + \Bigl[ \overline{C} - (3 D \!-\! 8) C_4\Bigr] a^{D-4} H^2 
F_{\mu\nu} F^{\mu\nu} + C_4 a^{D-6} \partial_{\alpha} F_{\mu\nu} \partial^{\alpha}
F^{\mu\nu} \; . \qquad \label{newcterms}
\end{eqnarray}
Note that we use the Minkowski metric to raise indices on the field strength
($F^{\mu\nu} \equiv \eta^{\mu\rho} \eta^{\sigma\nu} F_{\rho\sigma}$) and the
partial derivative operator ($\partial^{\alpha} \equiv \eta^{\alpha\beta}
\partial_{\beta}$).

From expression (\ref{newcterms}) we can read off how the coefficient of 
each counterterm affects the structure functions,
\begin{eqnarray}
\overline{C} - (3 D \!-\! 8) C_4 &\!\!\! = \!\!\!& +\frac12 \!\times\!
\frac{\kappa^2 \mu^{D-4}}{16 \pi^2 (D \!-\! 4)} \nonumber \\
& & \hspace{2cm} \Longrightarrow \Delta F_1 = -\frac{\kappa^2 H^2}{8\pi^2} 
\ln\Bigl( \frac{\mu a}{2 H} \Bigr) \delta^4(x \!-\! x') \; , \qquad 
\label{DF1} \\
C_4 &\!\!\! = \!\!\!& +\frac16 \!\times\! \frac{\kappa^2 \mu^{D-4}}{16 \pi^2 
(D \!-\! 4)} \nonumber \\
& & \hspace{0.5cm} \Longrightarrow \Delta F_2 = +\frac{\kappa^2}{24 \pi^2} 
\ln\Bigl( \frac{\mu a}{2 H}\Bigr)  \partial_{\mu} \frac1{a^2}
\partial^{\mu} \delta^4(x \!-\! x') \; , \qquad \label{DF2} \\ 
\Delta C - (D \!-\! 6) C_4 &\!\!\! = \!\!\!& -\frac23 \!\times\!
\frac{\kappa^2 \mu^{D-4}}{16 \pi^2 (D \!-\! 4)} \nonumber \\
& & \hspace{2cm} \Longrightarrow \Delta G 
= +\frac{\kappa^2 H^2}{6\pi^2} 
\ln\Bigl( \frac{\mu a}{2 H}\Bigr)  \delta^4(x \!-\! x') \; . \qquad 
\label{DG} 
\end{eqnarray}
Comparison with (\ref{Fimp}-\ref{Gimp}) reveals that neither (\ref{DF2}) nor
(\ref{DG}) is responsible for the factors of $\ln(a)$ in 
(\ref{photon}-\ref{Coulomb}). The factors of $\ln(a)$ all come from (\ref{DF1}),
which can be regarded as the coefficient of a curvature-dependent field strength
renormalization,
\begin{equation}
\delta Z \equiv -4 \Bigl[ \overline{C} - (3 D \!-\! 8) C_4\Bigr] H^2 = -
\frac{\kappa^2 H^2}{8 \pi^2} \!\times\! \frac{\mu^{D-4}}{D \!-\! 4} + 
O(\kappa^4 H^4) \; . \label{Zdef}
\end{equation}
The associated gamma function is,
\begin{equation}
\gamma \equiv \frac{\partial \ln(1 \!+\! \delta Z)}{\partial \ln(\mu^2)} =
-\frac{\kappa^2 H^2}{16 \pi^2} + O(\kappa^4 H^4) \; . \label{gamma}
\end{equation}

The Callan-Symanzik equation for $n$-point Green's functions is,\footnote{Change
$+n \gamma$ to $- n \gamma$ for one-particle-irreducible $n$-point functions.}
\begin{equation}
\Bigl[ \frac{\partial}{\partial \ln(\mu)} + \beta_{\kappa^2} 
\frac{\partial}{\partial \kappa^2} + n \gamma \Bigr] G_n\Bigl(x_1 ; x_2 ; 
\ldots ; x_n ; \mu ; \kappa^2\Bigr) = 0 \; . \label{CSeqn}
\end{equation}
The beta function for this theory goes like $\beta_{\kappa^2} \sim \kappa^4 H^2$,
so it does not affect 1-loop results. As one can see from (\ref{Fdef}-\ref{Gdef}),
the factors of $\ln(\mu)$ are always associated with factors $\ln(a)$ in the form
$\ln(\mu a)$. This is because primitive divergences produce no $D$-dependent scale 
factors, whereas the counterterms which absorb them not only contain a factor of 
$\mu^{D-4}$ but also a factor of $a^{D-4}$,
\begin{equation}
\frac1{D \!-\! 4} - \frac{\mu^{D-4} a^{D-4}}{D \!-\! 4} = -\ln(\mu a) + 
O(D \!-\! 4) \; . \label{logs}
\end{equation}
Hence we can replace the derivative with respect to $\ln(\mu)$ in expression
(\ref{CSeqn}) with a derivative with respect to $\ln(a)$. If we then regard the 
photon field strength (\ref{photon}) and the Coulomb potential (\ref{Coulomb}) 
as 2-point Green's functions it will be seen that the Callan-Symanzik equation 
(\ref{CSeqn}), with gamma function (\ref{gamma}), explains the factors of $\ln(a)$
in both results.

\section{Search for A Stochastic Explanation}

The previous section demonstrated that the factors of $\ln(a)$ in the photon
field strength (\ref{photon}) and the Coulomb potential (\ref{Coulomb}) can be
explained using a variant of the Renormalization Group. The purpose of this 
section is to explain why there seems to be no compelling variant of the 
stochastic formalism which explains the factor of $\ln(Hr)$ in the Coulomb 
potential. We begin by noting the characteristics of the $\ln(Hr)$ term. In 
particular, it may not even count as a ``leading logarithm'' effect as the 
factors of $\ln(a)$ do. We then discuss the problems with developing a
compelling stochastic explanation for it.

\subsection{Peculiarities of the $\ln(Hr)$ Term}

We have already mentioned that the factor of $\ln(H r)$ in the Coulomb 
potential (\ref{Coulomb}) derives from the nonlocal part of the vacuum 
polarization on the second line of expression (\ref{Fimp}). This descends
from the ``tail'' part of graviton propagator \cite{Miao:2018bol}; that 
is, from the logarithm part of $i \Delta_A(x;x')$ visible in expression 
(\ref{D4propA}). Its origin from the finite, nonlocal part of the graviton 
propagator means that the factor of $\ln(Hr)$ is not explainable by the 
Renormalization Group. If it is to be understood as a ``large logarithm'' 
we must seek a stochastic explanation based on a curvature-dependent 
correction to the electromagnetic field equation, similar to the 
curvature-dependent effective potentials which served to explain many
of the large logarithms in nonlinear sigma models \cite{Miao:2021gic}.

Before searching for a stochastic explanation we should discuss whether or
not the factor of $\ln(H r)$ qualifies as a ``large logarithm'' which should 
appear in the leading logarithm approximation. Many perfectly valid loop 
corrections are not recovered in this approximation. One example is the 
fractional correction of $2 G/(3 \pi a^2 r^2)$ in the Coulomb potential
(\ref{Coulomb}). This is the de Sitter descendant of a well-known flat 
space correction which was discovered by Radkowski in 1970 \cite{Radkowski:1970}.
It has nothing to do with inflationary particle production and clearly does
not belong to the leading logarithm approximation.

Because the initial manifold has coordinate radius comparable to the Hubble 
length \cite{Tsamis:1993ub}, we do not have access to the regime of $H r \gg 1$.
Hence the factor of $\ln(Hr)$ can only become large for $H r \ll 1$. That looks
more like an ultraviolet effect than an infrared one. In the same sense, the
Radkowski correction only becomes significant for small $r$. On the other hand,
the two effects depend very differently on the physical separation length
$a(t) H r$,
\begin{equation}
{\rm Radkowski} \longrightarrow \Bigl[ \frac1{a(t) H r}\Bigr]^2 \qquad {\rm versus}
\qquad {\rm Inflation} \longrightarrow \ln\Bigl[ a(t) H r\Bigr] \; . 
\label{comparison}
\end{equation}
The Radkowski effect only becomes large when the physical separation is small,
and for $aHr\ll 1$ it overwhelms the logarithm contribution,
whereas the inflationary effect is large when the physical separation becomes 
enormous.

\subsection{Problems with a Stochastic Explanation}

To understand our problems in deriving a stochastic formulation of electrodynamics
it is good to contrast the Lagrangian of electromagnetism plus gravity,
\begin{equation}
\mathcal{L}_{\rm EMGR} = \frac{(R \!-\! 2 \Lambda) \sqrt{-g}}{16 \pi G} - \frac14
F_{\rho\sigma} F_{\mu\nu} g^{\rho\mu} g^{\sigma\nu} \sqrt{-g} \; , \label{EMGR}
\end{equation}
with the nonlinear sigma model \cite{Miao:2021gic,Woodard:2023rqo} for which a 
compelling stochastic formulation exists,
\begin{equation}
\mathcal{L}_{AB} = -\frac12 \partial_{\mu} A \partial_{\nu} A g^{\mu\nu} \sqrt{-g}
- \frac12 \Bigl( 1 \!+\! \frac{\lambda}{2} A\Bigr)^2 \partial_{\mu} B \partial_{\nu}
B g^{\mu\nu} \sqrt{-g} \; . \label{AB}
\end{equation}
Both theories involve two fields, one of which engenders large logarithms and the
other not,
\begin{eqnarray}
h_{\mu\nu} \longrightarrow \Bigl( {\rm Logs}\Bigr) \qquad & , & \qquad A
\longrightarrow \Bigl( {\rm Logs}\Bigr) \; , \label{gmnA} \\
A_{\mu} \longrightarrow \Bigl( {\rm No\ Logs}\Bigr) \qquad & , & \qquad B
\longrightarrow \Bigl( {\rm No\ Logs}\Bigr) \; . \label{AmB}
\end{eqnarray}
(The graviton field $h_{\mu\nu}$ is defined by conformally transforming the metric,
$g_{\mu\nu} \equiv a^2 (\eta_{\mu\nu} + \kappa h_{\mu\nu})$.) The stochastic 
formulation of the nonlinear sigma model (\ref{AB}) was derived by integrating out 
the ``No Logs'' field $B$ from the equation of the ``Logs'' field $A$ in the 
presence of a constant $A$ background, 
\begin{eqnarray}
\lefteqn{\frac{\delta S[A,B]}{\delta A} = \partial_{\mu} \Bigl[ \sqrt{-g} \,
g^{\mu\nu} \partial_{\nu} A\Bigr] - \frac{\lambda}{2} \Bigl(1 \!+\! \frac{\lambda}{2}
A\Bigr) \sqrt{-g} \, g^{\mu\nu} \partial_{\mu} B \partial_{\nu} B \; , } 
\label{dSdA} \\
& & \hspace{0cm} \longrightarrow \partial_{\mu} \Bigl[ \sqrt{-g} \, g^{\mu\nu} 
\partial_{\nu} A\Bigr] - \frac{\lambda}{2} \Bigl(1 \!+\! \frac{\lambda}{2} A\Bigr) 
\sqrt{-g} \, g^{\mu\nu} \!\times\! \frac{\partial_{\mu} \partial'_{\nu} i
\Delta_A(x;x') \vert_{x'=x}}{(1 \!+\! \frac{\lambda}{2} A)^2} \; , \qquad
\label{intout} \\
& & \hspace{0cm} \longrightarrow \partial_{\mu} \Bigl[ \sqrt{-g} \, g^{\mu\nu} 
\partial_{\nu} A\Bigr] + \frac{\frac{3 \lambda H^4}{16 \pi^2} \sqrt{-g}}{1 \!+\!
\frac{\lambda}{2} A} \; . \qquad \label{stochAeqn}
\end{eqnarray}
This is a scalar potential model with potential $V_{\rm eff}(A) = -\frac{3 H^4}{
8 \pi^2} \ln\vert1 + \frac{\lambda}{2} A\vert$ and it can be treated using 
Starobinsky's stochastic formalism \cite{Starobinsky:1986fx,Starobinsky:1994bd}. 
Doing so recovers large logarithms in 1-loop corrections to the scalar mode 
function and the exchange potential \cite{Miao:2021gic}, as well as 1-loop and 
2-loop contributions to the expectation value of $A$ \cite{Miao:2021gic,
Woodard:2023rqo}.

The analog of the reduction (\ref{dSdA}-\ref{stochAeqn}) for our model 
(\ref{EMGR}) would be to integrate out the ``No Logs'' photon field from the 
``Logs'' metric field equation,
\begin{equation}
\frac{16 \pi G}{\sqrt{-g}} \frac{\delta S_{\rm EMGR}}{\delta g^{\mu\nu}} =
R_{\mu\nu} - \frac12 g_{\mu\nu} R + g_{\mu\nu} \Lambda - 8 \pi G \Bigl[
\delta^{\alpha}_{~\mu} \delta^{\beta}_{~\nu} g^{\rho\sigma} - \frac14 g_{\mu\nu}
g^{\alpha\beta} g^{\rho\sigma}\Bigr] F_{\alpha\rho} F_{\beta\sigma} \; .
\label{hmneqn}
\end{equation}
This might describe large logarithms affecting the graviton field 
\cite{Wang:2015eaa}, but it cannot capture the large logarithms 
(\ref{photon}-\ref{Coulomb}) induced by the graviton in the photon field. A 
stochastic explanation of those logarithms would presumably derive from 
integrating out the graviton from the photon field equation,
\begin{equation}
\frac{\delta S_{\rm EMGR}}{\delta A_{\mu}} = \partial_{\nu} \Bigl[ \sqrt{-g} \,
g^{\nu\rho} g^{\mu\sigma} F_{\rho\sigma} \Bigr] \; . \label{Ameqn}
\end{equation}
In the nonlinear sigma model (\ref{AB}) this would be like integrating out the
$A$ field from the $B$ equation,
\begin{equation}
\frac{\delta S_{AB}}{\delta B} = \partial_{\mu} \Bigl[\Bigl(1 \!+\! 
\frac{\lambda}{2} A\Bigr)^2 \sqrt{-g} \, g^{\mu\nu} \partial_{\nu} B\Bigr] \; .
\label{Beqn}
\end{equation}
That is exactly what was {\it not} done. Nor was there any stochastic explanation
for the explicit 1-loop and 2-loop results which were obtained for the field $B$
\cite{Miao:2021gic}. These results were all explained using the Renormalization 
Group. Moreover, integrating out the metric field would result in an
electromagnetic equation that still has derivative interactions,
precluding the stochastic formalism from being applied directly.

It is nevertheless undeniable that the graviton infrared modes are hugely enhanced,
and one might try to apply a perturbative version of the stochastic approximation
without integrating out any fields. For scalar potential models this amounts to 
approximating the real part of the $A$-type scalar propagator in~(\ref{simpprop}) 
by the corresponding infrared stochastic sum,
\begin{eqnarray}
&&
\hspace{-1.5cm}
{\rm Re} \Bigl\{ i \Delta_A(x;x') \Bigr\}
	\longrightarrow S(x;x')
	\equiv
		\int\! \frac{d^{3}k}{(2\pi)^3} \, e^{i\vec{k} \cdot (\vec{x} - \vec{x}^{\,\prime})}
\label{S def}
\\
&& 
	\times
	\theta(\varepsilon H a \!-\! k)
	\theta(\varepsilon H a' \!-\! k)
	\theta(k \!-\! \delta H)
	U(\eta,k) U^*(\eta',k)
	\, ,
	\qquad
	\varepsilon,\delta \ll 1 \, ,
\nonumber 
\end{eqnarray}
where the Chernikov-Tagirov-Bunch-Davies mode funtion is \cite{Chernikov:1968zm,
Bunch:1978yq},
\begin{equation}
U(\eta,k) = \frac{H}{ \sqrt{2k^3} } \Bigl[ 1 + i k\eta \Bigr] e^{-ik\eta}
	\xrightarrow{ -k\eta \ll 1} \frac{H}{ \sqrt{2k^3} } \, .
\end{equation}
This implies that the late time limit of~(\ref{S def}) is,
\begin{equation}
 S(x;x') = \frac{H^2}{4\pi^2}\times \ln(A)
 \,,\qquad\quad
 A = {\rm min}[a,a']
 \,.\quad
\label{S evaluated}
\end{equation}
%
%
%
The imaginary part of the propagator descends from inverting kinetic operators
of the equation of motion and should be kept as is.
This approximation is known to capture the leading infrared logarithms in massless 
scalar potential
models to all loops~\cite{Tsamis:2005hd,Woodard:2005cw,Prokopec:2007ak}, 
while the variant of this approximation 
adapted for light massive scalars is known to capture leading~$H^2/m^2\!\gg\!1$
corrections to 2-loop order~\cite{Kamenshchik:2021tjh}.

Applying this approximation to the one-graviton-loop correction to electromagnetism 
on de Sitter first requires expanding the photon field equation~(\ref{Ameqn})
in powers of graviton fluctuations,
\begin{eqnarray}
&&
- \eta^{\mu[\rho} \eta^{\sigma]\nu} \partial_\nu F_{\rho\sigma}
	-
	\frac{\kappa}{2} V^{\mu\rho\nu\sigma \alpha \beta} 
		\partial_\nu \Bigl[ h_{\alpha\beta} F_{\rho \sigma} \Bigr]
\nonumber \\
&&	\hspace{2.5cm}
	-
	\frac{\kappa^2}{2} U^{\mu\rho\nu\sigma \alpha \beta\gamma\delta} \partial_\nu 
		\Bigl[ h_{\alpha\beta} h_{\gamma\delta} F_{\rho \sigma} \Bigr]
	+
	\mathcal{O}(\kappa^3)
		= J^\mu
		\, .
\label{eom}
\end{eqnarray}
Here the 3- and 4-point vertex
tensor structures are~\cite{Leonard:2013xsa},
\begin{eqnarray}
V^{\mu\rho\nu\sigma \alpha \beta} 
	\!\!\!&=&\!\!\!
	\eta^{\mu[\rho} \eta^{\sigma] \nu} \eta^{\alpha \beta} 
	+ 4 \eta^{\alpha)[\mu} \eta^{\nu][\rho} \eta^{\sigma](\beta}
	\, ,
\\
U^{\mu\rho\nu\sigma \alpha \beta\gamma\delta} 
	\!\!\!&=&\!\!\!
	\biggl[ 
	\frac{1}{4} \eta^{\alpha\beta} \eta^{\gamma\delta} 
	- \frac{1}{2} \eta^{\alpha (\gamma} \eta^{\delta) \beta} 
	\biggr]
	\eta^{\mu [\rho}  \eta^{\sigma]\nu} 
	+
	\eta^{\gamma)[\mu} \eta^{\nu][\rho} \eta^{\sigma](\delta}\eta^{\alpha\beta} 
\nonumber \\
&&	\hspace{-2.4cm}
 	+ \eta^{\alpha)[\mu} \eta^{\nu][\rho} \eta^{\sigma](\beta} \eta^{\gamma\delta} 
	+ 
		\eta^{\mu (\alpha} \eta^{\beta) [\rho} \eta^{\sigma] (\gamma } \eta^{\delta) \nu}
		+
		\eta^{\mu (\gamma} \eta^{\delta) [\rho} \eta^{\sigma] (\alpha } \eta^{\beta) \nu}
	+
	\eta^{\mu [\rho} \eta^{\sigma] (\alpha} \eta^{\beta) (\gamma} \eta^{\delta) \nu} 
\nonumber \\
&&	\hspace{-2.4cm}
	+
	\eta^{\mu [\rho} \eta^{\sigma] (\gamma} \eta^{\delta) (\alpha} \eta^{\beta) \nu}
	+
	\eta^{\mu (\delta} \eta^{\gamma) (\alpha} \eta^{\beta) [\rho} \eta^{\sigma] \nu} 
	+
	 \eta^{\mu (\alpha} \eta^{\beta) (\gamma} \eta^{\delta) [\rho} \eta^{\sigma] \nu} 
	 \, .
\end{eqnarray}
We subsequently look for a perturbative solution of the field strength,
\begin{equation}
F_{\mu\nu} 
	= 
	F_{\mu\nu}^{(0)} 
	+ \kappa F_{\mu\nu}^{(1)} 
	+ \kappa^2 F_{\mu\nu}^{(2)} 
	+ \mathcal{O}(\kappa^3) \, .
\end{equation}
This is done by iterating the equation~(\ref{eom}) to 
order~$\kappa^2$,~\footnote{Note that in~(\ref{F2 eq}) we have not included 
the contribution formally of the same order descending from the Einstein 
equation~(\ref{hmneqn}). This contribution corresponds to the gravitational 
response to the photon, and does not harbor any large logarithms.}
\begin{eqnarray}
&& \hspace{-0.7cm}
	- \eta^{\mu[\rho} \eta^{\sigma]\nu} \partial_\nu F_{\rho\sigma}^{ (2)}(x)
	=
	\frac{\kappa^2}{2} U^{\mu\rho\nu\sigma \alpha \beta\gamma\delta} \partial_\nu 
		\Bigl[ \langle h_{\alpha\beta}(x) h_{\gamma\delta}(x) \rangle
			F_{\rho \sigma}^{ (0)}(x) \Bigr]
\label{F2 eq}
\\
&&	\hspace{0.cm}
	- \frac{\kappa^2 }{2} 
		\partial_\nu \biggl\{
		\int\! d^4x' \, \partial'_{\sigma} \partial'_\lambda G(x;x')
	V^{\mu\rho\nu\sigma \alpha \beta} \eta_{\rho\kappa}
	V^{\kappa\theta\lambda\phi \gamma\delta} 
		\langle h_{\alpha\beta}(x) h_{\gamma\delta}(x') \rangle
		F_{\theta\phi}^{ (0)}(x') 
		\biggr\}
		\, ,
\nonumber 
\end{eqnarray}
where the inverse of the flat space d'Alembertian~$\partial^2 \!=\! - \partial_0^2 \!+\! \nabla^2$ is,
\begin{equation}
G(x;x') = - \frac{\theta(\Delta\eta)}{4\pi} \frac{\delta (\Delta\eta \!-\! \| \Delta\vec{x} \| )}{ \| \Delta\vec{x} \| } 
	\, .
\end{equation}
The stochastic approximation then affects the graviton 2-point function~(\ref{simpprop}),
where the only contributing part is the one containing the~$A$-type propagator,
\begin{equation}
\langle h_{\mu\nu}(x) h_{\rho\sigma}(x') \rangle
	\longrightarrow
	\Bigl[ 2 \overline{\eta}_{\mu(\rho} \overline{\eta}_{\sigma)\nu}
		- 2 \overline{\eta}_{\mu\nu} \overline{\eta}_{\rho\sigma} \Bigr] S(x;x')
		\, ,
\end{equation}
where the stochastic sum~$S(x;x')$ is defined in~(\ref{S def}). Applying this prescription 
to the plane wave photon and to the Coulomb potential gives the following
contributions,
\begin{equation}
F^{0i}_{(2)} = F^{0i}_{(0)} \times \frac{\kappa^2 H^2}{2\pi^2} \ln(a) \, ,
\qquad \quad
\Phi_{(2)} = \Phi_{(0)} \times \frac{\kappa^2 H^2}{2\pi^2} \ln(a) \, .
\label{stoch corr}
\end{equation}
in the limit~$\varepsilon\!\ll\!1$. These contributions descend only from the
first term on the right-hand-side of Eq.~(\ref{F2 eq}), while the remaining nonlocal
term provides no leading order contributions. Not only does the Coulomb potential 
contribution in~(\ref{stoch corr}) fail to capture the~$\ln(Hr)$ term, but both
contributions overestimate the $\ln(a)$ corrections~(\ref{photon}--\ref{Coulomb}) 
from the full computation, that are completely captured by the RG explanation of 
Sec.~3. Upon closer examination, this discrepancy can be attributed to the lack 
of control over the cutoff parameter~$\varepsilon$. While for scalar potential 
models taking the limit~$\varepsilon\!\ll\!1$ remarkably works out to capture the 
leading contributions, in theories with derivative interactions this is not 
so,\footnote{Another example is 1-scalar loop corrections to the photon wave
function of scalar quantum electrodynamics \cite{Prokopec:2002uw,Prokopec:2003tm}.}
and the Hubble scale modes contribute relevant corrections, that for the system 
at hand have to cancel the contributions in~(\ref{stoch corr}).

The issue with applying the stochastic sum approximation to the graviton 
propagator is ultimately tied to derivative interactions, that are ubiquitous
in gravity. The issues arising from derivative interactions are well illustrated
by the mixed second derivative of the coincident propagator. The dimensionally 
regulated computation gives,
\begin{equation}
\langle \partial_\mu \phi(x) \partial_\nu \phi(x) \rangle
	= 
	-
	\frac{H^D}{ (4\pi)^{\frac{D}{2}} } \frac{ \Gamma(D) }{ 2 \, \Gamma(\frac{D+2}{2}) }
	g_{\mu\nu}
	\xrightarrow{D \to 4} 
	- \frac{ 3 H^4}{ 32 \pi^2 } 
	g_{\mu\nu}
	\,.
\label{coincident propagator: double derivative}
\end{equation}
However, when we apply the stochastic sum truncation to this
quantity one finds,
\begin{equation}
\langle \partial_\mu \phi(x) \partial_\nu \phi(x) \rangle
		\xrightarrow{a\to\infty} \frac{H^4}{8\pi^2}
	\biggl[
	\frac12a^2\delta_\mu^0 \delta_\nu^0\varepsilon^4
	+
	\frac{1}{3} \overline{g}_{\mu\nu}\varepsilon^2
	\biggr]
	\, , \label{stochastic}
\end{equation}
where $ \overline{g}_{\mu\nu} =  {g}_{\mu\nu}+a^2\delta_\mu^0 \delta_\nu^0$.
Whereas the exact result (\ref{coincident propagator: double derivative}) 
has a negative definite $\mu = i$, $\nu = j$ component, any stochastic 
mode sum such as (\ref{stochastic}) must produce positive definite results 
for the squares of operators. Derivative interactions prevent the affected
fields from carrying infrared logarithms, in which case these fields make
nonzero contributions of order one such as (\ref{coincident propagator: double derivative})
that come as much from the ultraviolet as from the infrared. No stochastic
mode sum can correctly describe these effects.

Another signal of problems in expression (\ref{stochastic}) is its strong
dependence on the cutoff. This arises in the stochastic formalism when the 
approximate scale invariance of the super-Hubble modes is either not present 
at tree level, or is suppressed by derivative interactions. For example, 
applying the stochastic formalism to vector fields in axion inflation results 
in a truncation which is sensitive to the cutoff~\cite{Talebian:2022jkb,
Fujita:2022fit}. Capturing large logarithms in these cases requires a 
systematic approach such as~\cite{Tsamis:2005hd,Woodard:2005cw,Prokopec:2007ak,
Vennin:2015hra,Moss:2016uix,Miao:2021gic,Woodard:2023rqo,Litos:2023nvj}.


\section{Conclusions}

The continuous production of gravitons during inflation is responsible for
the tensor power spectrum \cite{Starobinsky:1979ty} and for secondary effects
involving interactions with themselves and other particles. In chronological
order there have so far been six secondary, 1-loop effects reported on de 
Sitter background:
\begin{itemize}
\item{Enhancement of the fermion field strength \cite{Miao:2006gj};}
\item{Growth of the Coulomb potential in space and time \cite{Glavan:2013jca};}
\item{Enhancement of the photon field strength \cite{Wang:2014tza};}
\item{Enhancement of the graviton field strength \cite{Tan:2021lza};}
\item{Spatial suppression of the massless, minimally coupled scalar exchange 
potential \cite{Glavan:2021adm}; and}
\item{Suppression of the Newtonian potential \cite{Tan:2022xpn}.}
\end{itemize}
Prior to this work only the penultimate result had been given a Renormalization 
Group interpretation analogous to the stochastic-RG synthesis that was recently 
developed for nonlinear sigma models \cite{Miao:2021gic}. The terrific advantage 
of such an interpretation is that it permits an all-orders re-summation the 
series of leading logarithms. So it is wonderful news that we have here been 
able to provide a Renormalization Group explanation for the factors of $\ln(a)$ 
discovered in 1-graviton loop corrections to the Coulomb potential 
(\ref{Coulomb}) and the photon field strength (\ref{photon}). This was done in
section 3.

We were not able to achieve a similar explanation for the factor of $\ln(Hr)$ 
in the Coulomb potential (\ref{Coulomb}). Because this term derives from the 
nonlocal part of the vacuum polarization (see the second line of equation 
expression (\ref{Fimp}) for the structure function $F(x;x')$) the $\ln(Hr)$
does not appear to be associated with the mass scale $\mu$, the way the scale
factor $a(t)$ is through relation (\ref{logs}). In section 4 we searched for
a compelling stochastic explanation for the factor of $\ln(H r)$. We concluded
that none exists. The successful stochastic formulation of nonlinear sigma 
models \cite{Miao:2021gic} was derived by integrating out the derivative 
interactions (\ref{dSdA}-\ref{stochAeqn}), whereas it is the vector potential 
which is differentiated in the electromagnetic field equation (\ref{Ameqn}). 
Derivative interactions resist a stochastic interpretation because they 
mediate order one effects which derive from all parts of the dimensionally
regulated mode sum, rather than just from the leading infrared part. On the
other hand, we cannot integrate the vector potential out of its own equation
(\ref{Ameqn}), both because we want the resulting equation to describe 
electromagnetic effects and because the equation is linear in the vector
potential. We suspect that the lack of an explanation for the factor of 
$\ln(Hr)$ may indicate that it should not be considered a leading logarithm 
effect.

In nonlinear sigma models, which show both stochastic and RG effects 
\cite{Miao:2021gic,Woodard:2023rqo,Litos:2023nvj}, there are really three
things going on:
\begin{itemize}
\item{The generation of curvature-dependent, effective forces by integrating 
out differentiated fields in the presence of an approximately constant 
background;}
\item{The generation of stochastic jitter in the approximately constant 
background by the continual redshift of sub-horizon modes to the super-horizon;
and}
\item{The generation of secular logarithms through the incomplete cancellation
(\ref{logs}) between curvature-dependent primitive divergences and counterterms.}
\end{itemize}
As noted above, the first of these receives contributions from both ultraviolet 
and infrared, whereas the second is a purely infrared effect. We lump them
both under the rubric of ``stochastic'' because the second cannot occur without
the first, and we note again that there is no mechanism for producing the first
thing in the present analysis. The third thing does happen in our analysis and 
it is driven by the combination of ultraviolet electromagnetic modes with the
ultraviolet ``tail'' part of graviton modes.

The next step in our program is to attempt a similar explanation for the three 
remaining 1-graviton loop enhancements: the growing fermion field strength 
\cite{Miao:2006gj}, and the effects on gravitational radiation \cite{Tan:2021lza} 
and on the force of gravity \cite{Tan:2022xpn}. We anticipate that the fermionic 
effect will have a Renormalization Group explanation, as did the electromagnetic 
effects we considered here. However, explaining the two gravitational results 
may well require a stochastic analysis. That is as it should be because the
graviton field is analogous to the single field $\Phi$ in the nonlinear sigma
model analysis, and the factors of $\ln(a)$ in its mode function, exchange 
potential and expectation value all had a stochastic origin \cite{Miao:2021gic}.

Another step in our program is deriving the beta function $\beta_{\kappa^2}
\equiv \mu \frac{\partial \delta \kappa^2}{\partial \mu}$ so that we can use 
the Renormalization Group to derive all-orders results. This requires the 
portion of $\delta \kappa^2$ determined by a single loop of photons. A single 
matter loop of any sort induces two gravitational counterterms 
\cite{tHooft:1974toh,Barvinsky:1985an},
\begin{equation}
\Delta \mathcal{L}_{\rm GR} = c_1 R^2 \sqrt{-g} + c_2 C^{\alpha\beta\gamma\delta}
C_{\alpha\beta\gamma\delta} \sqrt{-g} \; , \label{GRcterms}
\end{equation}
where $R$ is the Ricci scalar and $C_{\alpha\beta\gamma\delta}$ is the Weyl 
tensor. The counterterm proportional to $c_2$ makes a higher derivative 
contribution term of no relevance to leading inflationary logarithms, however, 
the counterterm proportional to $c_1$ can be rewritten so that it contains a
part proportional to the Einstein-Hilbert Lagrangian,
\begin{equation}
R^2 = \Bigl[\!R - D (D\!-\!1) H^2\!\Bigr]^2 \!\!+ 2 D (D\!-\!1) H^2 \Bigl[\! 
R - (D \!-\! 1) (D \!-\! 2) H^2 \! \Bigr] + D (D\!-\!1)^2 (D\!-\!4) H^4 . 
\label{Eddington}
\end{equation}
Just as we regarded the middle term of (\ref{newcterms}) as a curvature-dependent
field strength renormalization so too we can think of the middle term of
(\ref{Eddington}) as a curvature-dependent renormalization of Newton's constant,
\begin{equation}
\delta \kappa^2 = -2 D (D \!-\! 1) c_1 \kappa^4 H^2 \; . \label{deltakappa}
\end{equation}
Because the factors of $\ln(\mu)$ are associated with $\ln(a)$ according to
relation (\ref{logs}), it should be noted that physical significance of our
beta function differs from the usual sense in which a negative sign means that
the theory becomes perturbative at high energy scales. For us it is the {\it 
positive} sign which betokens a perturbative theory at late times.

A final point is that this analysis has been made in the context of the simplest
graviton gauge \cite{Tsamis:1992xa,Woodard:2004ut}. We did not resolve the gauge
problem, nor must we do so in order to explain the large logarithms generated 
within a single gauge. Of course we should eventually employ the procedure for 
purging gauge dependence \cite{Miao:2017feh,Katuwal:2020rkv} to establish that 
the large logarithms are real, and to fix their numerical coefficients. Work on
this is far advanced \cite{Glavan:2021adm,Glavan:2023} but analyses in quantum 
gravity are so difficult that it is best to report on one at a time.

\vspace{.5cm}

\centerline{\bf Acknowledgements}

DG was
supported by the European Union and the Czech Ministry of Education, Youth and
Sports (Project: MSCA Fellowship CZ FZU I --- CZ.02.01.01/00/22\_010/0002906).
 SPM was supported by Taiwan NSTC grants 111-2112-M-006-038
and 112-2112-M-006-017. TP was supported by the D-ITP consortium, a 
program of the Neth\-erlands Organization for Scientific Research (NWO) 
that is funded by the Dutch Ministry of Education, Culture and Science 
(OCW). RPW was supported by NSF grant PHY-2207514 and by the Institute 
for Fundamental Theory at the University of Florida.

\end{document}